# Asynchronous Wi-Fi Control Interface (AWCI) Using Socket IO Technology

[1]Devipriya.T.K [2]Jovita Franci.A [3]Dr R Deepa [4]Godwin Sam Josh
[1,2]Student [3]Professor [4]Technical Director
[1,2,3]Department of Information Technology
[1,2,3]Loyola-ICAM College of Engineering and Technology Chennai, India [4]GloriaTech
Chennai, India

## Abstract

The Internet of Things (IoT) is a system of interrelated computing devices to the Internet that are provided with unique identifiers which has the ability to transfer data over a network without requiring human-to- human or human-to- computer interaction. Raspberry pi-3 a popular, cheap, small and powerful computer with built in Wi-Fi can be used to make any devices smart by connecting to that particular device and embedding the required software to Raspberry pi-3 and connect it to Internet. It is difficult to install a full Linux OS inside a small devices like light switch so in that case to connect to a Wi-Fi connection a model was proposed known as Asynchronous Wi-Fi Control Interface (AWCI) which is a simple Wi-Fi connectivity software for a Debian compatible Linux OS). The objective of this paper is to make the interactive user interface for Wi-Fi connection in Raspberry Pi touch display by providing live updates using Socket IO technology. The Socket IO technology enables real-time bidirectional communication between client and server. Asynchronous Wi-Fi Control Interface (AWCI) is compatible with every platform, browser or device.
**Keyword- Internet of Things, Web socket, Socket IO, Asynchronous Wi-Fi, Debian, Node.js, Raspberry Pi 3**

___

## I. INTRODUCTION

Internet of Things is the vast network of devices connected to the Internet, including smart phones and tablets and almost anything with a sensor on it and it includes cars, machines in production plants, jet engines, oil drills, wearable devices, and more. These "things" connected to the Internet called "Internet of Things" (IoT) and it collects and exchange data.

IoT is about making the data to come together in new ways for a business domain. The edge of the IoT is where the action is and it includes everyday physical objects that are becoming equipped with a wide array of sensors, actuators, and devices which are being (uniquely) addressable and interconnected, those system end-points that interact with and communicate real-time data from smart products and services. Technologies such as wireless sensor networks, short-range wireless communications, and radio-frequency identification (RFID) have allowed the Internet to penetrate in embedded computing. As IoT is growing fast, International Data Corporation (IDC) made a survey and found that by 2020 more than 28 billion things will be connected to the Internet from smart watches and other wearable's to smart cities, smart homes, and smart cars [1].

Raspberry pi-3 is small sized computer with Quad-core, 64-bit ARM and in-built WiFi module can be used in IoT projects to for establishing communication between the devices and to make them as "smart" devices [2], [3]. Raspberry pi can be attached to a small touch screen tablets available in the market to make it use in IoT projects. Most of the Raspberry pi is installed with Raspbian OS (Debian compatible Linux OS) and it is configured according to the project in which it is going to be used and when it is switched on, it opens directly with the specified configured software. In this case to connect to network connection Raspberry pi needs a Wi-Fi module which is proposed in this paper [4].

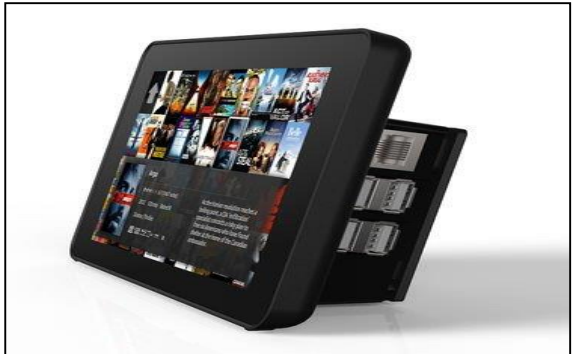

Fig. 1: Raspberry pi-3 with touch screen

Asynchronous Wi-Fi Control Interface (AWCI) scans the available Wi-Fi network continuously and updates the changes automatically by using Socket IO technology which makes use of Web socket protocol and this makes the system more live. Socket IO helps in developing a real time web application and it is a JavaScript library that enables real time, bidirectional communication between web clients and servers. The server side is developed using Node.js which is an event-driven architecture with a capability to support asynchronous I/O. In Web applications these design choices aim to optimize throughput and scalability with many input/output operations, as well as for real-time Web applications [6].

## II. ASYNCHRONOUS WI-FI CONTROL INTERFACE

AWCI is a touch screen optimized app to connect Raspberry-pi 3 to a wireless network. Devices that configure Raspbian OS for a special purpose and directly opens up with the specified software needs AWCI software component to connect wireless network. AWCI app will be more comfortable for any user to operate the device and easily make it connect to internet.

## III. SYSTEM ARCHITECTURE

User who needs to work on IoT projects with the help of Raspberry pi-3 and want their Raspberry pi-3 with a touch screen display wants to be connected to the wireless network can make use of AWCI software (touch screen optimized app) to connect to wireless network provided by a Wi-Fi Router by scanning available networks and by connecting to it when needed and disconnecting from the wireless network.

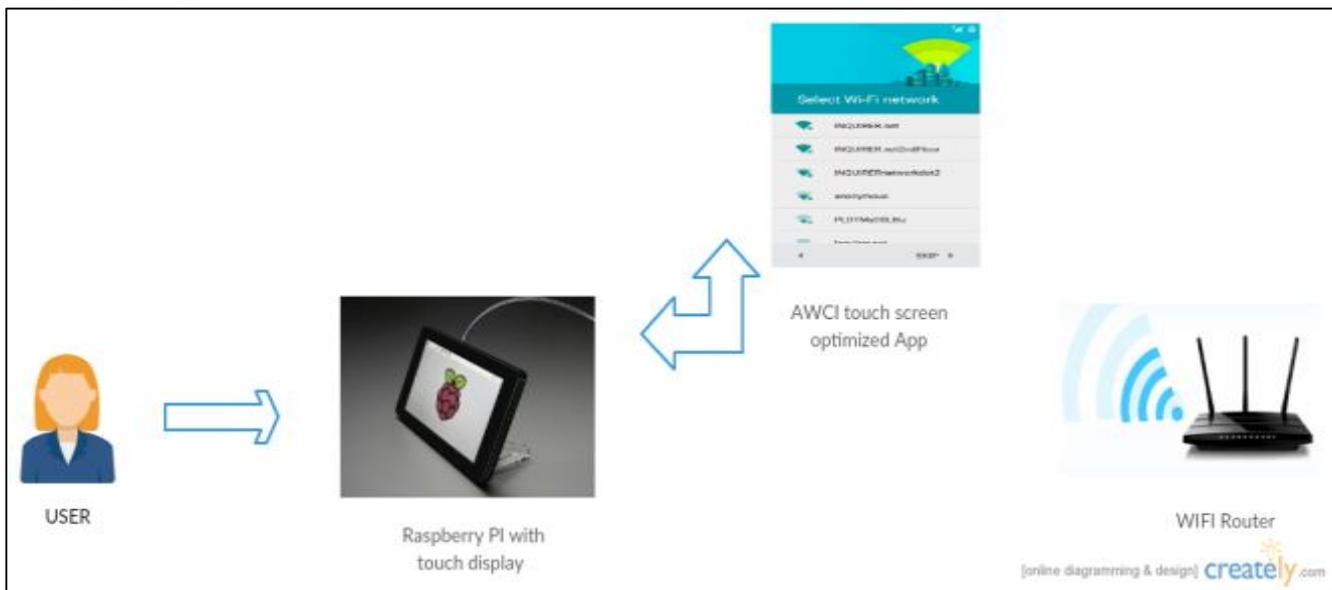

Fig. 2: System Architecture of AWCI

*A. Server Design Using Node.Js*

For developing a diverse variety of server tools and applications, Node.js which is an open-source, cross-platform JavaScript runtime environment can be used. In AWCI the Server side application is build using Node.js Although Node.js is not a JavaScript framework, many of its basic modules are written in JavaScript, which helps one to easily use this Node.js in developing their server application. Node.js applications can run on macOS, Microsoft Windows, and Unix servers (ie) it supports cross platform and so AWCI is a cross platform software. Node.js has an event-driven architecture capable of asynchronous I/O makes it has a choice to design Web applications with many input/output operations, as well as for real-time Web applications [5]. In AWCI Node.js was chosen as Server side as it supports asynchronous I/O which helps one to scan available networks continuously without refreshing the networks available page each time when a new network is available. Node.js have a collection of "modules" that handle various core functionality and these modules use an API designed to reduce the complexity of writing server applications.

Express is a flexible Node.js web application framework that provides a robust set of features to develop web and mobile applications. It facilitates the rapid development of Node based Web applications. It allows in setting up middleware to respond to HTTP Requests, defines a routing table which is used to perform different actions based on HTTP Method and URL, allows to dynamically rendering HTML Pages based on passing arguments to templates [7].

Socket IO library included in the Node.js helps to develop a real time bidirectional event-based communication. Initial step is setting up Node.js server and router, and then connecting socket.io to this server. After that sending data to the client through the socket connection and sending data to the server through the socket connection.

A Node.js module providing methods for scanning local WiFi access points, as well as connecting/disconnecting to networks and it works for Windows, Linux and MacOS. Among many available npm packages for WiFi connectivity "wireless-tools" is chosen in this paper as it is one among the other packages supported by Raspberry pi-3. This "wireless-tools" package supports "iwlist" which is command that is used to get detailed information from a wireless interface. It provides both synchronous and asynchronous methods. Along with "wireless-tools" another dependency "pi-wifi" is added to provide secure network connection via wpa-supplicant commands [8].

*B. Client Design Using Bootstrap*

Bootstrap is a free and open-source front-end web framework for designing websites and web applications. It contains HTML and CSS based design templates for typography, forms, buttons, navigation and other interface components, as well as optional JavaScript extensions. Bootstrap is compatible with the Google Chrome, Firefox, Internet Explorer, Opera, and Safari browsers, although some of these browsers are not supported on all platforms and this feature makes AWCI multi-platform software. It also supports responsive web design which means the layout of web pages adjusts dynamically, taking into account the characteristics of the device used (desktop, tablet, mobile phone) and this makes AWCI to dynamically adjust to the size of touch screen to which Raspberry pi-3 is connected [9],[10].

*C. Implementation of AWCI*

To connect to Internet, all the modules in AWCI are developed in such a way it is very similar to Android mobile Wi-Fi connection. Now a days since Android mobile phones are used by most of the users and it provide a very user friendly user interface similar user interface is used in AWCI.

*1) Scan Available Network Module*

This module scans available network and it display the available network with its strength and security when it is secured by a password. It displays the SSID (Service Set IDentifier) of the available networks. SSID is the name assigned to Wi-Fi network. All devices in the network must use this SSID (case-sensitive name) to communicate over Wi-Fi, which is a text string up to 32 bytes long. Wireless routers and access points have a default SSID, which may be the manufacturer's name. Some devices use their model number as the SSID. Using a Web browser, the SSID can be manually changed in the device's configuration settings. Wi-Fi routers normally broadcast their SSID and this helps AWCI to scan the broadcast SSID [11]. "Wireless-tools" asynchronous method "iwlist.scan ('wlan0', function (err, networks)" is used to scan the available network and return the scan results continuously with the help of call-back function.

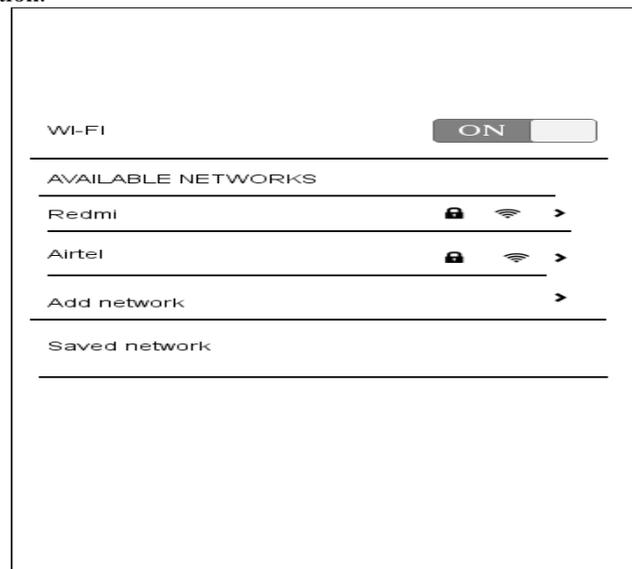

Fig. 3: Scans and display SSID of available network

*2) Connect to Network Module*

To connect to the available scanned network one can click to the particular SSID. If it is secure it will display a modal dialog box to enter the password and it will validate the password and if it is valid it allows one to connect to a wireless network else if it not valid it display a "Password Incorrect" message only authenticated user can be able to connect to wireless network in case password is provided. If it is not secured with a password it will connect directly to that particular SSID and one will be connected to a wireless network. Once it is connected it displays "Connected" message indicating that connection is established. "wireless-tools" asynchronous method "wpa_supplicant.enable(options, function(err)" is used to connect to the scanned network and this use a call-back to report continuous updates like authenticating, connecting, and so on.

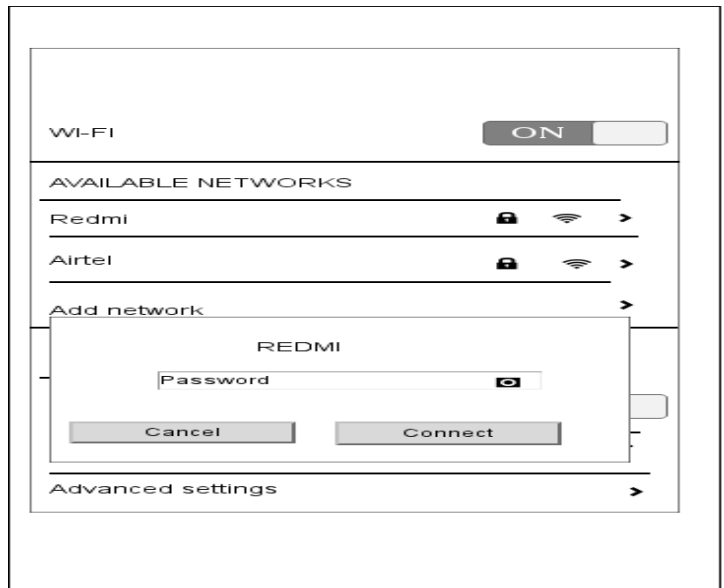
Fig. 4: Provide password for a particular SSID (REDMI) to connect to network

*3) Disconnect From Network Module*

To disconnect from the connected wireless network one should again click the connected SSID which have a "Disconnect" button clicking that one can disconnect from the wireless network connection. "wireless-tools" asynchronous method "wpa_supplicant.disable('wlan0', function(err)" is used to disconnect from the connected network and sometimes powering down and back up can take a while, depending on user's wireless card, so this method will return callback when it is complete.

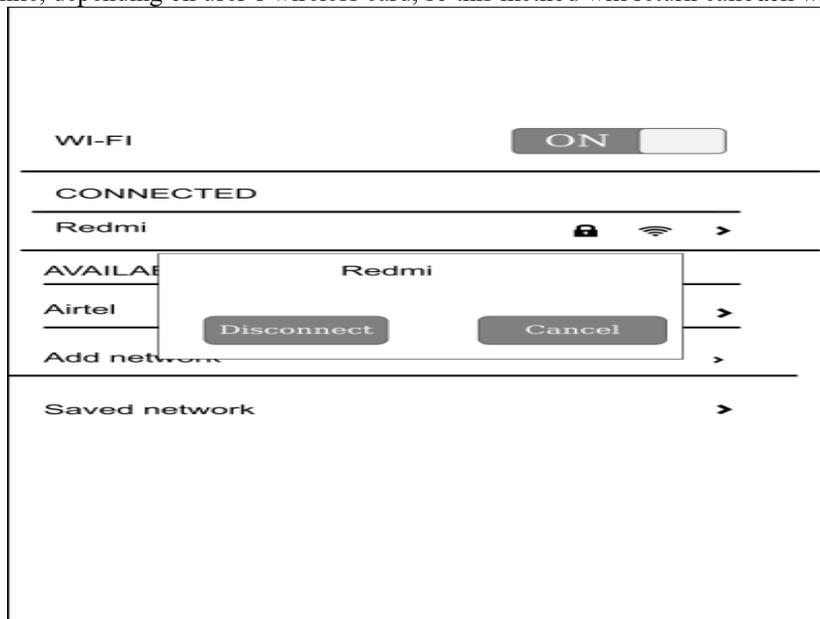
Fig. 5: Click disconnect button to disconnect from wireless network

## IV. CONCLUSION AND FUTURE ENHANCEMENT

In this paper, a simple Wi-Fi connectivity software Asynchronous Wi-Fi Control Interface (AWCI) was developed for Raspberry pi-3 which has in built Wi-Fi and it can be used in Internet of Things (IoT) projects to connect to Internet and make their devices "smart" with the help of Raspberry pi-3 with a touch screen optimized.

Further enhancement can be done by developing another Socket IO client who can view the available networks of another client remotely and connect to the available networks remotely using Socket IO technology since it has the capability to establish connection even in the case of personal firewall and antivirus software which prevents external device in entering the internal private network. This can be useful in case an organization develops a Wi-Fi connection for its client but the client found some issues in connecting to available networks [12].